\documentclass{interact}

\usepackage{titlesec}
\usepackage{graphicx}
\usepackage[version=3]{mhchem} 
\usepackage{graphicx}
\usepackage{subcaption}
\usepackage{array,multirow}
\usepackage{amsmath}
\usepackage{todonotes}
\usepackage{hyperref}
\usepackage{soul}
\usepackage{algpseudocode,algorithm,algorithmicx}
\usepackage[utf8x]{inputenc}

\usepackage{fancyhdr}

\begin{document}

\title{Cross entropy as objective function for music generative models.}

\author{
\name{Sebastian Garcia-Valencia\textsuperscript{a,b}\thanks{Correspondence: Sebastian Garcia Valencia, Computer Science Department, Universidad EAFIT Carrera 49 No 7 Sur-50 Medellin, Colombia. Email: sgarci18@eafit.edu.co}}
\affil{\textsuperscript{a}Computer Science Department, Universidad EAFIT, Medellin, Colombia; \textsuperscript{b}Research and Development Department, AIVA Technologies, Luxembourg City, Luxembourg}
}

\maketitle

\begin{abstract}
The election of the function to optimize when training a machine learning model is very important since this is which lets the model learn. It is not trivial since there are many options, each for different purposes. In the case of sequence generation of text, cross entropy is a common option because of its capability to quantify the predictive behavior of the model. In this paper, we test the validity of cross entropy for a music generator model with an experiment that aims to correlate improvements in the loss value with the reduction of randomness and the ability to keep consistent melodies. We also analyze the relationship between these two aspects which respectively relate to short and long term memory and how they behave and are learned differently.
\end{abstract}

\begin{keywords}
Music; Recurrent Neural Network; Consistency; Memory; NLP
\end{keywords}

\section{Introduction} \label{sec:introduction}

For any machine learning model, the election of an objective function is an important part of the process. It is this function that will let us know how well the outputs of the model are when compared with the real data. The decision is not trivial, many criteria like the shape and mechanics of the function are important features to have into account. The first one will help to avoid falling into local minima that will lead the model to get stuck in the training just by choosing a function with a more concave tendency. The second will be influenced by the type of model and problem.

In the case ML models, the first question to ask is if the problem is a regression or a classification problem, in the first case, a function like the mean square error will fit well since in this case, the output will be a single real number while in classification the result is typically an array of all the possible outputs with percentages of how sure is the model about each of them, making them a probability distribution.

The generative models of sequences are pretty much a classification problem since you topically will say which element of the dictionary of elements is the next, you are classifying each input element and then using the output as input in the next step, metrics like log-likelihood and techniques like evaluation based in nearest neighbors have been explored in the past \cite{2015arXiv151101844T}. Music generation can be seen as a sequence generation problem where these elements of the sequences are notes. Different approaches have been used to calculate this loss, using reinforcement learning with metrics based in music theory to define the rewards the magenta project  \cite{DBLP:journals/corr/JaquesGTE16} reduced unpleasant characteristics like notes out of key.

Another common strategy is to use a defined metric as objective function in a classic optimization problem. Usually, cross-entropy is a popular choice for classification and generative models \cite{Johnson2017, Hutchings2017, whorley2016, kawthekarevaluating}. According to information theory, the cross-entropy explicitly calculates how different two probability distributions are \cite{shannon48}. 

Cross entropy is usually denoted by formula \ref{eq:cross1} where p(x) is the original probability distribution (in this case our ground truth) and q(x) is the predicted distribution (the output of the model). 

\begin{equation}
    H(p,q)=-\sum_{x_i} p(x_i)\log q(x_i)
    \label{eq:cross1}
\end{equation}

Formula \ref{eq:cross2} gives a better intuition on how it works in a machine learning classification problem. When the probability value in $q(x_i)$ is too low (the model doesn't believe this is the right answer), the logarithm will make the value very negative what will make the second term of the summation smaller. The highest the probability in $q(x_i)$ the bigger the value of the term. The value of $p(x_i)$, will let us ignore all the values in wrong predictions since $p(x_i)$ will be 0 in this cases, discarding that value.  We will want to optimize this value since at the end it will mean "make the probability of the right answer higher"

\begin{equation}
    H(p,q)=\sum_{x_i} p(x_i)\log \frac{1}{q(x_i)}
    \label{eq:cross2}
\end{equation}

These features make cross entropy  an ideal option for classification since the output of your model is a probability distribution. The metric will quantify the prediction capability \cite{kawthekarevaluating}. Being the ground truth an unknown distribution, it is necessary to use and estimator for it, which in this case is the negative average log-likelihood. Commonly, in NLP, the metric chosen is the perplexity, which is the exponentiation of the cross-entropy. For optimization purposes both are equivalent, however, the use of perplexity brings the metric to a linear scale what makes it easier to interpret.

The purpose of this experiment is to validate if the cross entropy, extensively used for text in NLP also works for music sequences, we want to check how the improvement in the cross entropy relates musically to the output melodies, specifically if it progressively shows reductions of randomness and better consistency end to end. Finally, we want to know if these two features, related to short and long term memory respectively develop differently.

The remaining part of this paper is organized as follows: Chapter \ref{sec:experiments} presents the experimental design. Chapter \ref{sec:results} analyses the cross entropy evolution and make a musical analysis of the melodies generated in each case. Finally, chapter \ref{sec:conlusions}, summarizes the discoveries.

\section{Experiments} \label{sec:experiments}
A recurrent neural network based in LSTM cells was trained using cross entropy as objective function, this means that the output of the last layer was compared with the actual y value (the actual next note in the sequence), since the output of the last layer is a distribution array telling us the probability of each position to be the right prediction, it is comparing how similar is this to the real distribution. Then it is passed to an ADAM optimizer which calculates the loss and changes the weights of the neural network using backpropagation.

We will save the weights of the model and check the value of the loss at different points of the training, namely at iteration 0 (the beginning where the weights are just random numbers), after 20, 30, 750, and 40000 (the final weights). Using these weights we will use the sampling mode of the model to generate a melody using always the same seed, the sequence of notes C4-D4-E4-D4, then we ask the model to generate 30 notes\footnote{You can listen how this evolution sounds in \url{https://youtu.be/sIUu5mYU0l8}}.

\section{Results} \label{sec:results}

This section shows how the improvement in cost function value, corresponds with a better structure and reduction of randomness in the generated melody.

\begin{figure}[h]
    \centering    
    \includegraphics[width=\textwidth]{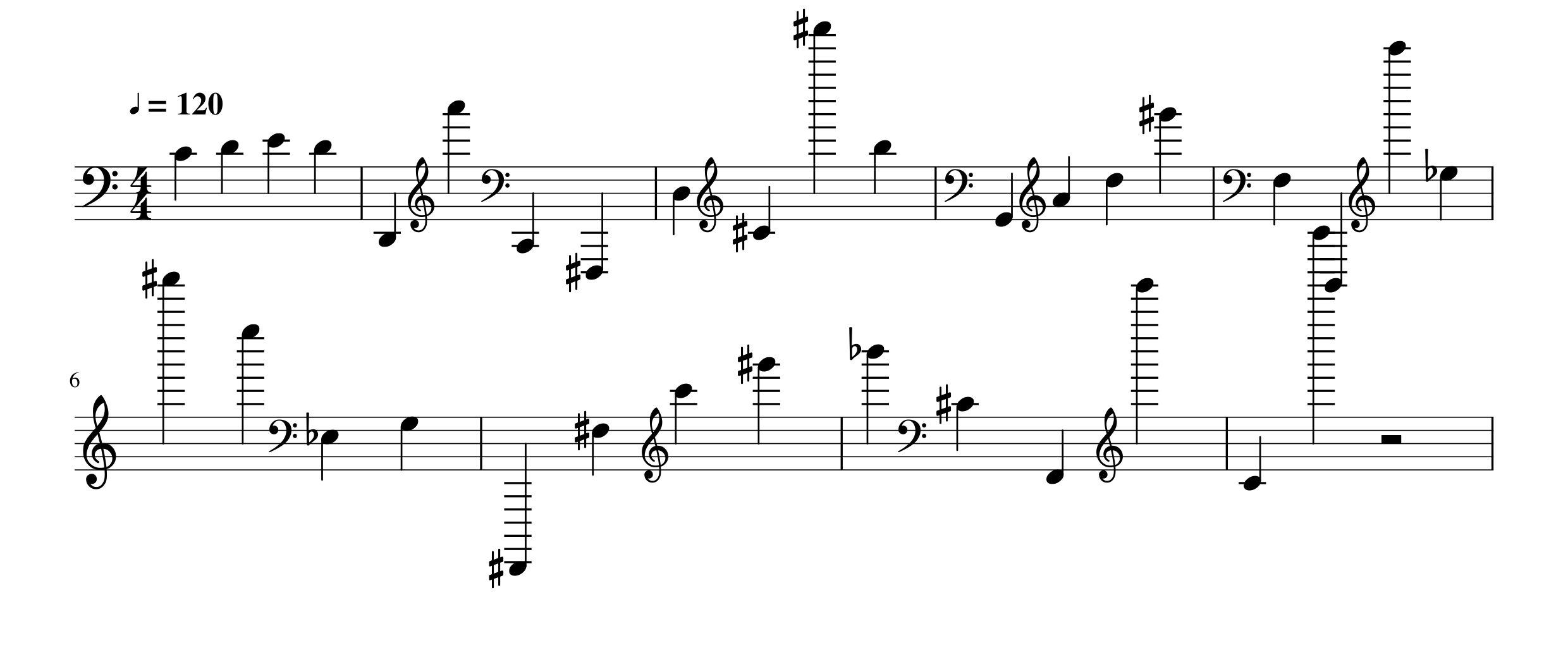}
    \caption{Step 0 loss 4.68086}
    \label{fig:costfuncmelody1}
\end{figure}

At the beginning (fig \ref{fig:costfuncmelody1}) the model has a cross entropy of 4.68, it is evident that the generated notes are just the result of random selection, the notes changes produce big intervals that make no sense and even go further than the limits of the visualization software. 

\begin{figure}[h]
    \centering    
    \includegraphics[width=\textwidth]{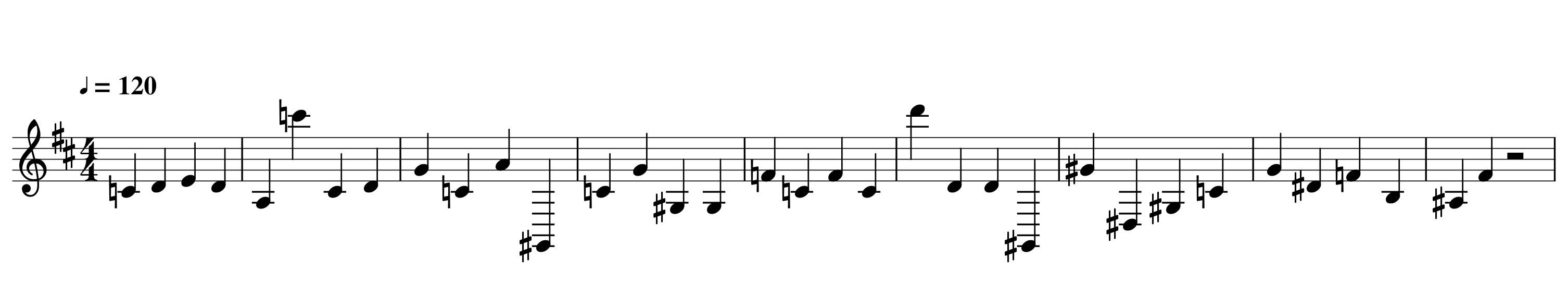}
    \caption{Step 20 loss 3.74101}
    \label{fig:costfuncmelody2}
\end{figure}

After only 20 iterations and a loss of 3.74 (fig \ref{fig:costfuncmelody2}), the model reduces the randomness, but the interval of the changes is still significant, the main problem is that it still does not sound like a song, the motif that begins with C$\sharp$ on the third beat of bar 2 and continues with D-G-C-A, makes you thing is getting an idea, especially because of the perfect fifth below (G-C) and the subsequent minor sixth at the end, but the jump to the G$\sharp$ two octaves lower kill the flow. 

\begin{figure}[h]
    \centering    
    \includegraphics[width=\textwidth]{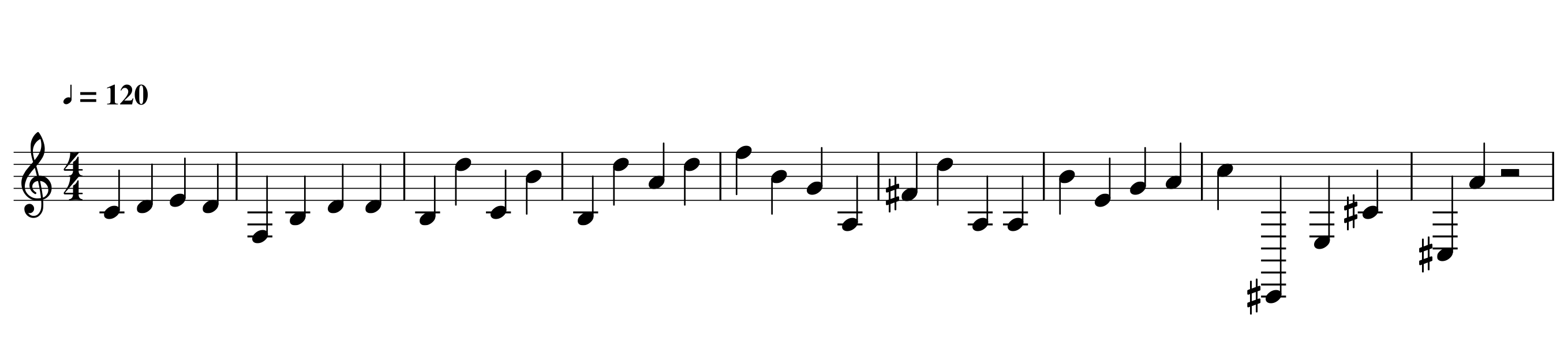}
    \caption{Step 30 loss 3.19856}
    \label{fig:costfuncmelody3}
\end{figure}

10 iterations later, with 3.19 of loss (fig \ref{fig:costfuncmelody3}), it is easy to notice the apparition of some basic patterns. There are basically 3 motifs, one in bars 2-3, one in bars 4-7 and one in bars 8-9. They are relatively consistent internally, with some exceptions like the last A in bar 5 which is still evidence of randomness. Also, the 3 motifs are disconnected between them, with interval jumps too big for a melody.

\begin{figure}[h]
    \centering    
    \includegraphics[width=\textwidth]{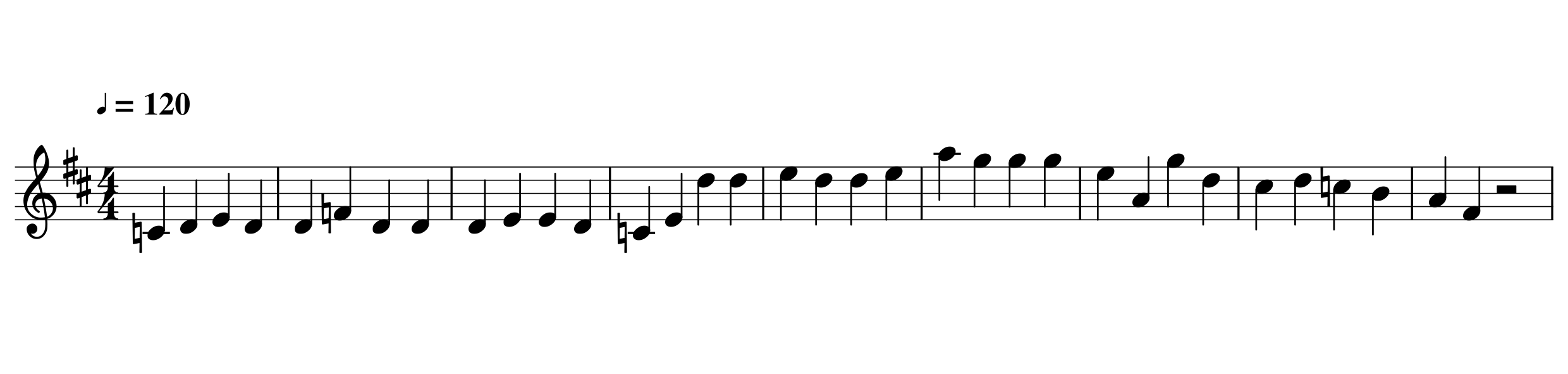}
    \caption{Step 750 loss 2.19148}
    \label{fig:costfuncmelody4}
\end{figure}

After 750 iterations the randomness disappears, but the model still cannot keep the melody consistent from end to end, the change from the second to the third note in bar 4(fig \ref{fig:costfuncmelody4}) is evidence of this, it goes from E4 to D5 and feels like beginning a completely different melody, this tells us that the neural network still has not developed enough memory context to keep consistency for a longer period of time, however, is remarkable the minor second below (D5-C$\sharp$5) at the end of bar 7 and the minor third below (A4-F$\sharp$4) at the end, which far from randomness, give the piece a flavor of a typical modulation in classical music.
 
\begin{figure}[h]
    \centering    
    \includegraphics[width=\textwidth]{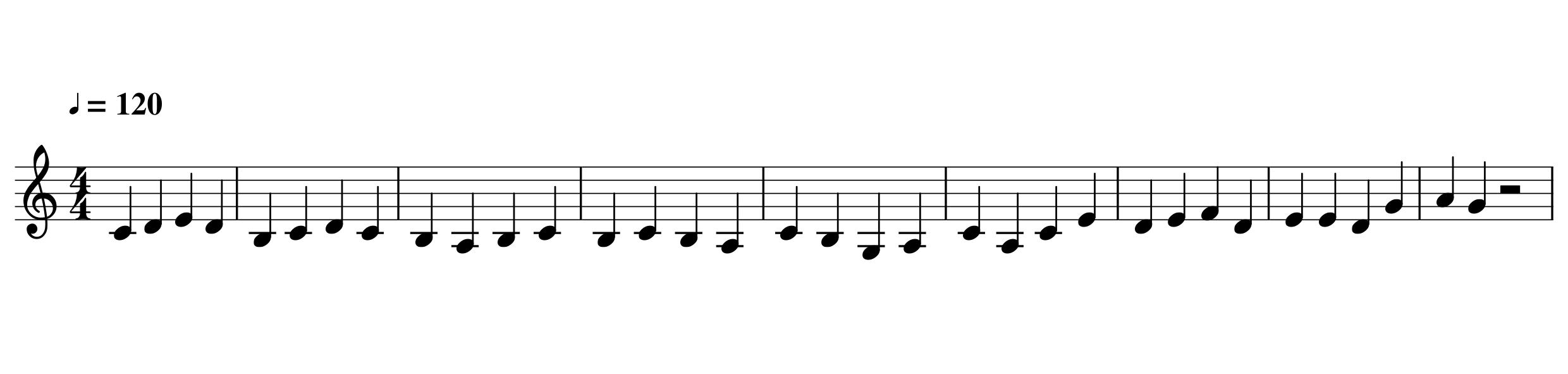}
    \caption{Step 40000 loss 1.59703}
    \label{fig:costfuncmelody5}
\end{figure}

Finally, after 40000 iterations and reaching a loss of 1.59 (fig \ref{fig:costfuncmelody5}), the result is a complete melody which keeps the fluidity from start to end. The biggest interval found in this generated melody is the perfect fourth (D4-G4) at the end of bar 8 which by no means cut the flow of the melody, the rest of the intervals are seconds or thirds, with only 2 major thirds, one below in bar 5 (B3-G3) and one in bar 6 (C4-E4). This makes the melody predominantly made with the 3 shortest intervals, which makes it even suitable as a piece for voice.

\begin{figure}[h]
    \centering    
    \includegraphics[width=\textwidth]{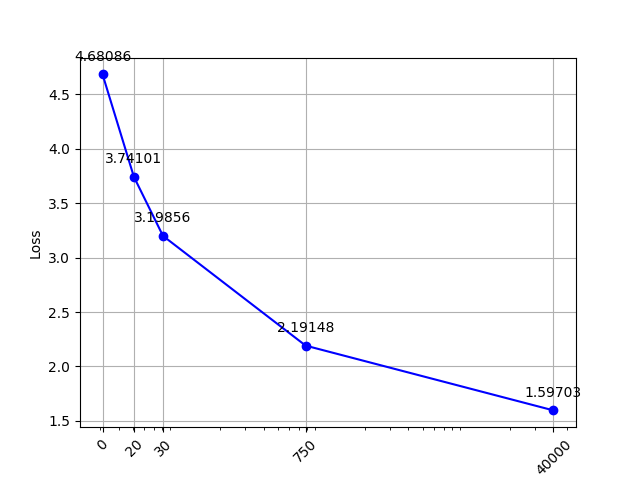}
    \caption{Loss value in each iteration}
    \label{fig:losschart}
\end{figure}

It is interesting how there is a significant improve in the melodies related to the reduction in loss from 4.68 to 2.19 in just the first 750 iterations (Fig. \ref{fig:losschart}) but it takes 39250 iterations more to reduce it to 1.59 and reach the point where the model is able to keep consistency in the melody along the 30 notes.

\section{Conclusions} \label{sec:conlusions}
The evidence found in section \ref{sec:results} shows that cross entropy works perfectly fine as the objective function for a machine learning generative model when the sequences are melodies. Just as it happens with text generation, where the optimization of the loss based in cross entropy correlates with the development of meaning and consistency in the text produced, the perceived tonality, randomness, and consistency of the melodies improve in direct correlation with the metric.

Respect to the last two features, it is important to mention that reduction in randomness can be reached relatively fast in early iterations, however, the development of the song takes many more training iterations, from this we can conclude that this type of sequence model can learn a local context very fast (Fig. \ref{fig:costfuncmelody1}, \ref{fig:costfuncmelody2} and \ref{fig:costfuncmelody3}), getting some intuition of the surrounding elements that make sense for a given note, while the more long term context, crucial for the consistency across all the melody, involves the development of deeper structures to have the context well established in the memory cells, which is only possible after a high enough number of repetitions (Fig \ref{fig:costfuncmelody4} and \ref{fig:costfuncmelody5}).

It is also remarkable how the long term consistency relates to the reduction in the loss, as Fig. \ref{fig:losschart} shows, it took just the 2\% of the time to gain the 80.5\% of the total improvement (from 4.68 to 2.19 in 750 iterations), and then it takes the remaining 98.12\% of the iterations for the remaining 19.5 (from 2.19 to 1.59 in 39250 iterations), however, is in this small improvement where the model reaches the whole consistency of the melody.

\bibliographystyle{apalike}
\bibliography{bibliography.bib}

\end{document}